\pdfoutput=1
\documentclass[fleqn,usenatbib]{mnras}

\usepackage{newtxtext,newtxmath}

\usepackage[T1]{fontenc}
\usepackage{ae,aecompl}

\usepackage{graphicx}	
\graphicspath{{./Figures/}}
\usepackage{amsmath}	
\usepackage[caption=false]{subfig}

\title[Adiabatic compression of relic in A3411-3412]{The application of the adiabatic compression scenario to the radio relic in the galaxy cluster Abell 3411-3412}

\author[C. Button \& P. Marchegiani]{
Charissa Button$^{1,2}$\thanks{E-mail: charissa.button1@students.wits.ac.za}
and Paolo Marchegiani$^{1,3}$
\\
$^{1}$School of Physics, University of the Witwatersrand, 1 Jan Smuts Avenue, 2000, South Africa\\
$^{2}$Department of Physics, University of Pretoria, Lynwood Road, 0002, South Africa\\
$^{3}$Dipartimento di Fisica, Universit\`{a} La Sapienza, P. le A. Moro 2, Roma, Italy
}

\date{Accepted XXX. Received YYY; in original form ZZZ}

\pubyear{2020}

\begin{document}
\label{firstpage}
\pagerange{\pageref{firstpage}--\pageref{lastpage}}
\maketitle

\begin{abstract}
Although radio relics are understood to originate in intracluster shock waves resulting from merger shocks, the most widely used model for describing this (re-)acceleration process at shock fronts, the diffusive shock acceleration (DSA) model, has several challenges, including the fact that it is inefficient at low shock Mach numbers. In light of these challenges, it is worthwhile to consider alternative mechanisms. One possibility is the adiabatic compression by a shock wave of a residual fossil electron population which has been left over from a radio galaxy jet. This paper applies this model to the relic hosted in the merging galaxy cluster Abell 3411-3412, where a radio bridge between the relic and a radio galaxy has been observed, with the aim to reproduce the spatial structure of the spectral index of the relic. Four scenarios are presented, in which different effects are investigated, such as effects behind the shock front and different shock strengths. The results show that the adiabatic compression model can reproduce the observed spectral indices across the relic for a shock Mach number that is lower than the value required by the DSA-type modelling of this relic and is in accordance with the values derived from X-ray observations, if other mechanisms, such as an expansion phase or post-shock turbulence, are effective behind the shock front.
\end{abstract}

\begin{keywords}
galaxies: clusters: general -- galaxies: clusters: individual: A3411-3412 -- radio continuum: galaxies
\end{keywords}



\section{Introduction}
In addition to the radio emission due to the cluster galaxies, diffuse radio emission is also present in galaxy clusters. This emission is produced through synchrotron radiation as relativistic electrons (with Lorentz factors of $\gamma \gg 1000$ and particle number densities of $\sim 10^{-10} \text{ cm}^{-3}$) move in helical paths through large scale magnetic fields which have intensities of $\sim 0.1 - 10 \text{ } \muup \text{G}$ \citep[e.g.][]{Feretti_et_al_2012, Brunetti_&_Jones_2014, van_Weeren_et_al_2019}. The detection of these diffuse sources indicates that non-thermal components, such as large scale magnetic fields and populations of relativistic particles, are mixed into the thermal plasma of the intracluster medium (ICM). Diffuse radio emission is usually present in clusters where shock fronts have been observed in X-rays and microwave through the Sunyaev-Zel’dovich effect \citep[e.g.][]{Markevitch_&_Vikhlinin_2007, Erler_et_al_2015}, or in clusters that have a high level of turbulence \citep{Eckert_et_al_2017}. Because of this, the detection of diffuse radio emission can be used to select clusters in a non-relaxed state, or also as a complementary method to discover clusters that were missed in other bands \citep[e.g.][]{Macario_et_al_2014, de_Gasperin_et_al_2017}.

This diffuse emission has been divided into three main categories: radio haloes, radio relics, and radio mini-haloes. Both radio haloes and radio relics are extended ($\gtrsim 1$ Mpc), low-surface brightness ($\sim 0.1 - 1 \text{ } \muup \text{Jy} \text{ arcsec}^{-2}$ at 1.4 Ghz) sources found in merging clusters that have a steep radio spectral index ($\alpha \leq -1, F_{\nu} \propto \nu^{\alpha}$) and that do not have optical counter parts. Radio haloes tend to occur at the centre of galaxy clusters and have a regular morphology with little to no polarization, whereas radio relics tend to occur at the cluster periphery and have either a roundish or an elongated morphology and are strongly polarized ($20 - 30 \%$). Radio mini-haloes are similar to radio haloes but are less extended ($\lesssim 500$ kpc) and are typically found around the brightest cluster galaxy in cool-core clusters \citep[see e.g.][for a review]{Feretti_et_al_2012}.

A long-standing problem in the study of radio relics is the origin of the relativistic electron population. In the Bohm description of the diffusion of cosmic rays in a plasma with fluctuations in the magnetic field, the typical diffusion length-scale in the ICM of a GeV electron within its radiative lifetime ($\lesssim 10^8$ Gyr) is of the order of 10 pc \citep{Bagchi_et_al_2002}. Although plasma motions can increase this distance to some extent, it still remains less than 1 Mpc \citep{van_Weeren_et_al_2019}. Even in the case of Kolmogorov diffusion, the diffusion length-scale, which can be on the order of tens of kpc \citep{Brunetti_2003}, is still less than the typical extent of relics. Thus, for Mpc-size relics, the electron cooling time is much less than the possible time since the electrons could have been injected from a galaxy at the location of the relic. This requires that an efficient particle acceleration mechanism is present within or close to radio relics \citep{Ensslin_et_al_1998,Kang_et_al_2012}. \citet{Ensslin_et_al_1998} demonstrated that accretion or merger shock waves in the ICM could (re-)accelerate electrons within a magnetic field behind the shock front, indicating that relics trace shock waves. The clearest evidence of the connection between radio relics and shocks is provided by the detection in the X-rays of shock fronts co-located with radio relics \citep[e.g.][]{Akamatsu_&_Kawahara_2013}. Further evidence of this relationship between relics and shock waves is provided by the correlation between the merger axis of the cluster and the relic position, with the relic major axis usually being perpendicular to the merger axis, as well as the fact that the radio power of relics scales with the X-ray luminosity and the mass of the host cluster \citep{Vazza_et_al_2015}.

Although there is compelling evidence that the relativistic electron population is created through ICM shock waves, and possibly also turbulence, \mbox{(re-)ac}celerating the electrons to relativistic energies \citep{van_Weeren_et_al_2019}, there is still some debate as to the particle acceleration mechanism. A widely adopted mechanism is diffusive shock acceleration (DSA) which is based on a first order Fermi process where thermal electrons diffusing through the ICM can cross the shock front several times, each time gaining energy and eventually reaching relativistic energies \citep{Drury_1983, Blandford_&_Eichler_1987}.

However, this model has several challenges. While it is fairly efficient in accelerating thermal electrons at strong shocks, at weak shocks produced by galaxy cluster mergers it is inefficient \citep{Kang_et_al_2012}. The estimated acceleration efficiency of cosmic ray (CR) protons is less than 1\% \citep{Ryu_et_al_2019, Botteon_et_al_2020}, while, due to the higher rigidity of thermal protons compared to thermal electrons, the DSA model is expected to be even less efficient in the acceleration of CR electrons \citep[see also][]{Kang_et_al_2012}. \citet{Vazza_&_Bruggen_2014} showed that for about half of the relics in the sample that they investigated, the required electron acceleration efficiency is greater than the predicted value from the DSA model (see fig.~1 in their paper). It has also been found that in some conditions a shock with a Mach number $< \sqrt{5}$ might not be able to accelerate particles \citep{Vink_&_Yamazaki_2014}.

Furthermore, the CR protons, accelerated via DSA, can diffuse through the cluster over distances of several hundred kpc \citep[e.g.][]{Blasi_&_Colafrancesco_1999} and should then produce gamma-rays as a result of collisions with thermal protons. However, the predicted flux of these gamma-rays in nearby and rich clusters is at or above the upper limit set by \textit{Fermi} observations \citep[e.g.][]{Vazza_&_Bruggen_2014,Vazza_et_al_2015, Vazza_et_al_2016} and therefore requires a lower efficiency for DSA processes so that these limits are not violated.

Another challenge to the DSA model is an observed discrepancy between the value of the shock Mach number determined within the DSA model and that determined from X-ray observations. Within the DSA model the shock Mach number is given by
\begin{equation}
\mathcal{M}^2 = \frac{2 \alpha - 3}{2 \alpha + 1}
\text{,}
\label{eqn: Shock Mach number in DSA theory}
\end{equation}
\citep[e.g.][]{Blandford_&_Eichler_1987} where $\alpha$ is the radio spectral index and is related to the electron spectral index, $s$ where $N(\gamma) \propto \gamma^{-s}$, by $\alpha = - \left( s-1 \right)/2$. It has been found that the shock Mach number predicted by DSA is sometimes higher than the Mach number measured from X-ray observations, indicating that the DSA model at the X-ray Mach number is not sufficient to produce the observed radio spectral index \citep{Akamatsu_et_al_2017, Colafrancesco_et_al_2017, Botteon_et_al_2020}.

In order to deal with these issues, the standard DSA model acting on the thermal electron pool was modified to a DSA model acting on a fossil population of mildly relativistic electrons \citep{Pinzke_et_al_2013}. This idea is supported by observations where the relic is observed to be connected to a radio galaxy, which is a natural source for fossil relativistic plasma. Examples of clusters in which the relic is connected to a radio galaxy are the cluster, PLCKG287.0+32.9 \citep{Bonafede_et_al_2014}, the Bullet cluster \citep{Shimwell_et_al_2015}, and the cluster Abell 3411-3412 \citep{van_Weeren_et_al_2017}. While this would reduce the required efficiency of the shock, it would still not explain the lack of gamma-ray detection in galaxy clusters \citep{Vazza_&_Bruggen_2014}. This suggests that other models, not based on DSA, need to be considered as possible mechanisms for producing CR electrons in relics.

\citet{Ensslin_&_Gopal-Krishna_2001} proposed a model that is based on the adiabatic compression of clouds of fossil radio plasma rather than on the direct acceleration of electrons. In this model, a cloud of fossil radio plasma, previously emitted by a radio galaxy, is energized as a shock wave passes around the cloud compressing it adiabatically. The shock wave does not penetrate the cloud since the speed of sound within the plasma is greater than the speed of sound in the ambient medium. As the cloud of plasma is compressed, the electrons and magnetic field gain energy causing the cloud to become radio luminous. This model could explain the relative rareness of relics since it requires both a shock wave and fossil radio plasma that has not aged too much. It could also explain the trend that relics form at the periphery of clusters since the reviveable age of fossil plasma at the periphery is greater than that at the centre due to the stronger synchrotron losses at the centre. Like DSA models, the adiabatic model can explain the thin elongated morphology of relics, since the emission is confined to a narrow shell behind the shock front \citep[see also][]{Zhang_et_al_2019}. Furthermore, the observed level of polarization in radio relics can also be produced through compression \citep{Ensslin_&_Bruggen_2002}. On the other hand, the adiabatic compression model only works as long as the relativistic plasma remains well confined inside the cloud. When mixing with the surrounding ICM starts to take place, the sound speed inside the bubble can drop and the DSA mechanism can effectively reaccelerate particles, even though the Mach number inside the cloud is lower compared to that outside the cloud \citep{Kang_2018}.

The adiabatic compression mechanism as a possible mechanism for the origin of some radio relics has been considered by \citet{Kempner_et_al_2004} who classified the known relic sources as radio gischt and radio phoenices assuming a different origin mechanism for the two classes of sources. In this classification radio gischt are associated with the ICM and are produced by a shock wave accelerating electrons from the thermal plasma. On the other hand, radio phoenices are associated with extinct or dying AGN and are produced through the adiabatic compression of the fossil radio plasma that is left over from AGN. For this reason, these sources have been classified as revived fossil plasma sources by \citet{van_Weeren_et_al_2019}. The recent cases, listed above, where sources classified as radio gischt have been shown to be connected to radio galaxies challenge this classification of relics and could suggest that there is common origin for both sources classified as radio gischt and sources classified as radio phoenices \citep[see also][]{Colafrancesco_et_al_2017}.

This paper investigates whether the adiabatic compression model is a plausible model for the formation of the radio relic hosted in the galaxy cluster A3411-3412 by investigating whether it is able to reproduce the observed radio spectral shape across the relic. This relic has been observed to be connected to a nearby cluster galaxy through a narrow radio tail which can provide the fossil electron cloud that the adiabatic model requires. Although \citet{van_Weeren_et_al_2017} were able to reproduce the spectral properties of the relic using a DSA-type re-acceleration model, their modelling required a higher Mach number compared to the Mach number determined through the X-ray observations. This paper will investigate whether the adiabatic model can reproduce the radio flux density and spectral shape of the relic with a lower Mach number.

The structure of the paper is as follows. Section 2 reviews previous observations of the relic hosted in the cluster Abell 3411-3412 and discusses the previous attempt to model the relic using a DSA-type model. Section 3 outlines the adiabatic model and how it was used in this case to model the relic. Section 4 presents the results of the adiabatic compression model applied to this relic. Finally, section 5 presents the conclusions.

A flat $\Lambda$ cold dark matter cosmological model with $\Omega_{M} = 0.27$, $\Omega_{\Lambda} = 0.73$ and $H_{0} = 71 \text{ km} \text{ s}^{-1} \text{ Mpc}^{-1}$ is used throughout the paper.

\section{The relic in the galaxy cluster Abell 3411-3412}
\label{sec: The case of Abell 3411-3412}
The galaxy cluster Abell 3411-3412 (hereafter A3411-3412), located at a redshift of $z = 0.164$, is currently undergoing a merging event of the two sub-clusters, A3411 and A3412, where the merger axis lies along the SE-NW direction \citep{van_Weeren_et_al_2013}. The two merging clusters have similar masses ($\sim 10^{15} \text{ M}_{\odot}$ each) and are viewed about $1$ Gyr after the first core passage \citep{van_Weeren_et_al_2017}.

\citet{van_Weeren_et_al_2013} observed a region of diffuse radio emission to the south-east of the cluster centre along the major axis of the extended X-ray emission. They observed this region to be made up of five elongated components with a total extent of $1.9$ Mpc. From the size of the emission, its location with respect to the cluster centre, the lack of optical counterparts and its polarization properties, \citet{van_Weeren_et_al_2013} classify this region as a radio relic. \cite{van_Weeren_et_al_2017} found a cluster galaxy that is connected through a radio tail to the north-eastern portion of the relic as well as two more galaxies, one of which has a tail of steep spectrum emission extending towards the relic and the other is embedded within the relic emission, suggesting that all the components of the complex relic could trace revived fossil plasma.

From their X-ray observations, \citet{van_Weeren_et_al_2017} detected a shock front at the relic's outer edge. The jump in their X-ray surface brightness profile indicates that the shock is rather weak with a Mach number of $1.2 \lesssim \mathcal{M} \lesssim 1.7$. More recent X-ray observations indicate that the shock can be even weaker with a Mach number of only $\mathcal{M} =1.13$ with a 90\% confidence upper limit of $\mathcal{M}<1.6$ \citep{Andrade-Santos_et_al_2019}. However, these derivations of the shock Mach number are based on a broken power-law density model and the assumptions of the projection effects and shock geometry might be incorrect which could result in an underestimation of the shock Mach number. Because temperature measurements are less affected by an unknown geometry, a better method of determining the shock Mach number is to use the temperature discontinuity. However, this is often difficult in the external regions of clusters where a small number of counts in X-ray measures is expected. Thus, these values of the shock Mach number should be considered as representative values.

This work focuses on the north-eastern component of the relic (labelled R1 by \citet{van_Weeren_et_al_2013}). The radio tail connecting the relic to the cluster galaxy could be a source of radio plasma and the presence of a shock at the position of the relic provides a way for the fossil plasma to be re-energized by adiabatic compression. We note also that the size of the region R1 is of the order 250 kpc, i.e. smaller than a typical giant radio relic. This could suggest a peculiar origin for this region, making it interesting to explore adiabatic compression as the mechanism for its origin. If this mechanism is successful, this small region might be classified as re-energized AGN-plasma placed inside a wider radio relic.

\citet{van_Weeren_et_al_2017} measured the radio flux density at intervals of 20 kpc across the relic, starting from a point on the outer edge of the relic, as well as the spectral index at four points around the shock front (see fig.~9 of their paper). These measurements allow us to determine the flux density and the spectral index at the position of the shock front. Since the radio plasma is revived by the shock wave, the position with the greatest flux density should correspond to the position of the shock. In front of the shock, the plasma is still to be revived and so it should have a steeper spectrum and lower radio emission compared to the emission at the position of the shock, while the plasma behind the shock loses energy rapidly due to radiative emission and so would also have a steeper spectrum and lower radio emission. Based on their data, we can assume that the position of the shock front is between 80 and 100 kpc. It is interesting to note that the flux densities measured at the higher frequencies (1.5 and 3.0 GHz) reach a peak value at 100 kpc while the flux densities at the lower frequencies (325 and 610 MHz) reach a peak value at 80 kpc. This makes it difficult to determine the exact position of the shock front.

According to the results found by \citet{van_Weeren_et_al_2017}, the spectral index between 0.325 and 3.0 GHz of the radio tail connecting the relic to the cluster galaxy steepens from $\alpha = -0.5 \pm 0.1$ at the galaxy nucleus to $\alpha = -1.3 \pm 0.1$ at the end of the radio bridge. However, at the position where the radio plasma from the galaxy joins the relic, the spectral index flattens to $\alpha = -0.9 \pm 0.1$. The spectral index along the length of the relic remains largely flat, although it steepens across the relic towards the cluster centre. The same authors measured the spectral index between 0.325 and 3.0 GHz at four points across the relic.

\citet{van_Weeren_et_al_2017} applied a DSA-type re-acceleration model to this north-eastern component of the relic. Although their modelling is able to explain the uniform spectral index along this relic, it predicted a Mach number of 1.9 which is higher than the Mach number determined from their own X-ray observations. Furthermore, they also suggested that this DSA-type re-acceleration model could have difficulties explaining the constant spectral indices along the lengths of other, larger relics, unless the associated shock has a high Mach number or the spectral index distribution of the fossil plasma is fairly uniform. In light of the recent observational and theoretical challenges to DSA based models it is useful to investigate other models for the origin of this relic.

\section{The adiabatic compression model}
\subsection{Overview of the model}
\label{Sec: Formalism}
The adiabatic compression model was described by \citet{Ensslin_&_Gopal-Krishna_2001} and takes place in several phases, labelled phase 0 to phase 4. The initial phase, phase 0, describes the injection of radio plasma into an expanding cocoon by an active galaxy. During phase 1, the radio cocoon continues to expand after the radio galaxy becomes inactive, causing the electrons to lose energy adiabatically. In addition, the magnetic field is weakened by the expansion. This, and the adiabatic energy losses of the electrons, can result in the radio emission of the cocoon to become much fainter than the emission of phase 0. In phase 2, the radio plasma is in pressure equilibrium with the environment and the electrons exist at low energies as a result of the previous adiabatic losses. Additionally, the magnetic field is weakened due to the expanded state of the cocoon. Since the radiative losses depend quadratically on the particle energy and the magnetic field strength \citep{Sarazin_1999}, these losses are suppressed during this phase which enables the cocoon to maintain its energy. This fossil plasma is revived during phase 3, where the plasma is compressed during a shock passage caused by a merger event. Since the shock Mach number in the cocoon is smaller than the shock Mach number in the thermal medium due to the higher sound speed in the cocoon, the shock wave is not expected to penetrate the cocoon but rather to compress it. This compression causes the electron population and the magnetic field to gain energy adiabatically leading to a sharp increase in the synchrotron emissivity. During phase 4, the final phase of the model, the plasma is again in pressure equilibrium with the post-shock medium and the synchrotron emission of the relic fades away as a result of the heavy radiative losses.

The resulting electron spectrum at the end of each phase is calculated, according to the formalism presented in \citet{Ensslin_&_Gopal-Krishna_2001}, from the characteristic momentum of each phase. This characteristic momentum is specified by the compression ratio of the phase, $C_{i-1\,i}$, the expansion or compression exponent, $b_i$, the time-scale of expansion, $\tau_i$, and the duration of the phase, $\Delta t_i$. The compression ratio is defined as the ratio of the volume at the end of phase $(i-1)$ to the volume at the end of phase $i$, where it is assumed that the volume changes according to a power-law in time:
\begin{equation}
V(t) = V_0 \left( \frac{t}{t_0} \right)^b
\text{, or }
C(t) = \left( \frac{t}{t_0} \right)^{-b}
\text{,}
\label{eqn: Volume change as power law}
\end{equation}
where $b$ is the expansion or compression exponent. It follows, then, that for an expansion phase the ratio is less than $1$, while for a compression phase the ratio is greater than $1$. The exponent $b_i$ is related to the rate of the expansion or compression of the radio cocoon. The time-scale $\tau_i$ is a characteristic time-scale of the expansion or the compression which is also related to the rate of expansion or compression. Both the characteristic time-scale, $\tau$, and the exponent $b$ are determined by the physical process under which the volume of the plasma changes. Lastly, the quantity $\Delta t_i$ is the phase duration.

These four quantities are related by the equation
\begin{equation}
C_{i-1 \, i} = \left( 1 + \Delta t_i / \tau_i \right)^{-b_i}
\text{.}
\label{eqn: Phase parameters}
\end{equation}
Thus, in order to describe a phase in the model, three of the four parameters need to be determined and the remaining one can be calculated from the above equation.

From the resulting electron spectrum the luminosity and flux density can be calculated from standard synchrotron formulae \citep[e.g.][]{Schlickeiser_2002}.

\subsection{Determination of parameters}
\label{sec: Determination of parameters}
Some of the phase parameters during each phase can be estimated through physical considerations of the active galaxy, the radio plasma and the thermal environment. However, other parameters for some phases can only be fixed when applying the model to particular clusters and considering the cluster properties such as the shock strength and the pre-shock magnetic field. The parameters that are fixed within the model are described in detail in \citet{Ensslin_&_Gopal-Krishna_2001} and are listed in Table~\ref{tab: Phase parameter values}.

The compression ratio of phase 1, $C_{0\,1}$, has to be less than 1 to represent an expansion phase. If both the magnetic field of the radio galaxy, $u_{\text{B\,0}}$ and the pre-shock magnetic field $u_{\text{B\,2}}$ are known, $C_{0\,1}$ can be calculated since $C_{1\,2} = 1$ \citep[see][]{Ensslin_&_Gopal-Krishna_2001}. However, in this case the magnetic field of the galaxy is unknown, so the value of $C_{0\,1}$ had to be determined by fitting the data before and after this phase. Then, the phase duration, $\Delta t_1$, is determined by eqn \eqref{eqn: Phase parameters}. During phase 2, three of the four phase parameters are fixed, but since the expansion time-scale is infinite, the phase duration, $\Delta t_2$, is not determined by eqn \eqref{eqn: Phase parameters}. Instead, for this phase, the duration is a free parameter, but since the electron population does experience radiative losses, the phase cannot last too long so that the radio emission of the revived plasma is still detectable. The duration of phase 3 is related to the volume of the plasma at the end of phase 2 and the pre-shock velocity in the shock frame \citep{Ensslin_&_Gopal-Krishna_2001}. However, as it is difficult to estimate the volume of the radio plasma, it is easier to treat $\Delta t_3$ as a free parameter. The phase compression ratio, $C_{2\,3}$ is estimated from the pressure jump, $P_3/P_2$, of the surrounding thermal gas, assuming pressure equilibrium before and after the shock passage:
\begin{equation}
	C_{2\,3} = \left( \frac{P_3}{P_2} \right)^{3/4}
	\text{,}
	\label{eqn: Compression ratio phase 3}
\end{equation}
where the pressure jump in thermal gas is related to the shock Mach number:
\begin{equation}
	\frac{P_3}{P_2} = \frac{2 \gamma_{\text{B}} \mathcal{M}^2}{\gamma_{\text{B}} +1} - \frac{\gamma_{\text{B}} -1}{\gamma_{\text{B}} +1}
	\label{eqn: P_3/P_2 from the Mach no}
\end{equation}
\citep{Landau_&_Lifshitz_1959} assuming the behaviour of a perfect gas. Here $\gamma_{\text{B}}$ is the adiabatic index of the thermal gas: $\gamma_{\text{B}} = 5/3$. As in phase 2, the expansion time-scale of phase 4 is infinite so that the phase duration, $\Delta t_4$, is not fixed by eqn \eqref{eqn: Phase parameters} but is left as a free parameter that should be determined in each application of the model. Again, since the electron population is losing energy very rapidly due to radiative losses, the phase duration strongly affects the level of radio emission produced by the relic. The values or constraints of the phase parameters are listed in Table \ref{tab: Phase parameter values}.

\begin{table*}
	\caption[Summary of the phase parameters that are determined within the adiabatic model]{Summary of the values of each of the phase parameters. The values that are marked with a dash are determined from a fit to the observed cluster properties in each specific application.}
	\label{tab: Phase parameter values}
	\centering
	\begin{tabular}{ccccc}
		\hline
		\rule{0pt}{2.2ex}Phase & {$\Delta t_i$} & {$\tau_i$} & {$C_{i-1 \, i}$} & {$b_i$}
		\\
		& Myr & Myr & &
		\\
		\hline
		\rule{0pt}{2.2ex}{0} & 0 & 15.0 & 1 & 1.8
		\\
		{1} & $\left( (C_{0\,1})^{-1/b_1} - 1 \right)\tau_1$ & 10.0 & $<1$ & 1.2
		\\
		{2} & --- & $\infty$ & 1 & 0
		\\
		{3} & --- & $\left( (C_{2\,3})^{-1/b_3} - 1 \right)^{-1}\Delta t_3<0$ & $\left(\frac{P_3}{P_4}\right)^{3/4}>1$ & 2
		\\
		{4} & --- & $\infty$ & 1 & 0
		\\
		\hline
	\end{tabular}
\end{table*}

\subsection{Application of the model}
The remaining parameters have to be determined in the application of the model to the radio relic in A3411-3412. The compression ratio of phase 3 is determined from the value of the shock Mach number which was taken to be $\mathcal{M} = 1.7$, according to the upper limit obtained by \citet{van_Weeren_et_al_2017}. From this value, the compression ratio of phase 3 is calculated from eqns \eqref{eqn: Compression ratio phase 3} and \eqref{eqn: P_3/P_2 from the Mach no} to be $C_{2\,3} = 2.48$.

Since phase 3 of the adiabatic model describes the action of the shock front to revive the fossil plasma, the spectral index at the end of this phase should correspond to the spectral index at the position of the shock front. As discussed in section~\ref{sec: The case of Abell 3411-3412}, the exact position of the shock front is difficult to determine, but the flattest measured value of the spectral index is $-0.95$ and occurs at 80 kpc. This modelling, therefore, attempts to fit the spectral index at the end of phase 3 to this value. Of the values of the spectral index across the relic that are available, the value of $-1.59$ that occurs at 140 kpc is the one that is furthest from the shock front. Thus, the value of the spectral index at the end of phase 4 is matched to this value. Since the initial electron population is injected by the radio galaxy, the spectral index at phase 0 is matched to the value of the spectral index of the galaxy, $-0.5$. The electron cloud at the end of the radio tail corresponds to the electron population at the end of phase 2 before the action of the compression caused by the shock passage. Thus, the desired spectral index at the end of phase 2 corresponds to the spectral index at the point of conjunction between the radio tail and the radio relic which is $-1.3$  \citep[][]{van_Weeren_et_al_2017}. The value of the spectral index at the end of phase 1 is not constrained; however, due to the expansion during this phase it will be steeper than the spectral index at the end of phase 0. The expected spectral index for each phase is listed in Table \ref{tab: Target spectral index values.}. The values of the free parameters, $C_{0\,1}$, $\Delta t_2$, $\Delta t_3$, and $\Delta t_4$, were varied until the calculated spectral indices at the end of each phase matched these spectral indices across the relic. The values of $\Delta t_1$ and $\tau_3$ are calculated from the values of $C_{0\,1}$ and $\Delta t_3$ according to eqn \eqref{eqn: Phase parameters}.

\begin{table}
	\caption{Target values of the spectral index between 0.325 and 3.0 GHz at the end of each phase.}
	\label{tab: Target spectral index values.}
	\centering
	\begin{tabular}{cl}
		\hline
		\rule{0pt}{2.2ex} {Phase} & {Spectral index}
		\\
		\hline
		\rule{0pt}{2.2ex}0 & $-0.5$
		\\
		1 & Not specified
		\\
		2 & $-1.3$
		\\
		3 & $-0.95$
		\\
		4 & $-1.6$
		\\
		\hline
	\end{tabular}
\end{table}

The CMB radiation field energy density, $u_{\text{C}}$, is given by the blackbody radiation density where the temperature depends on the redshift of the galaxy cluster, according to the equation 
\begin{equation}
u_{\text{C}} = \frac{4 \sigma T^4}{c} = \frac{4 \sigma (2.725(1+z))^4}{c}
\text{.}
\label{eqn: CMB radiation field energy density}
\end{equation}
Thus, for the redshift of the cluster galaxy connected to the radio tail, $z = 0.164$, \citep{van_Weeren_et_al_2017}, the corresponding CMB radiation field energy density is $u_{\text{C}} = 7.66 \times 10^{-13} \text{ erg cm}^{-3}$ or $u_{\text{C}} = 0.478 \text{ eV cm}^{-3}$.

The value used for the pre-shock magnetic field strength was the same as that used in the modelling performed by \citet{van_Weeren_et_al_2017}: $B = 1.5 \text{ } \muup \text{G}$. In this model, the pre-shock magnetic field corresponds to the magnetic field of phase 2. The values of the magnetic field in the other phases were calculated from this value. 

The injection spectrum was found by estimating the electron spectrum of the galaxy. The radio spectrum of $-0.5$ is best reproduced by a power-law spectrum with a slope of $-2$:
\begin{equation}
N_0(p_0) = \tilde{N_0} p^{-2}
\text{.}
\label{eqn: Electron spectrum galaxy, phase 0}
\end{equation}
Since the magnetic field of the galaxy is unknown and its flux density is not reported in the literature, the constant $\tilde{N_0}$ was estimated a posteriori by matching the calculated flux density at the end of phase 3 with the peak flux density of the relic. The best-fitting value is $\tilde{N_0} = 2.2 \times 10^{61}$.

Using the values of these parameters, the evolution of the electron population was determined at the end of each phase of the model and the flux density at the end of each phase was calculated from these electron spectra.

\section{Results and discussion}
This section presents the results of four scenarios in which the adiabatic compression model was applied to the relic in A3411-3412, and discusses the results that were obtained. The first scenario investigates the application of the model according to the formalism presented above. In order to obtain a better fit to the data at the end of phase 4, scenarios two and three investigate different effects behind the shock front. Finally, the fourth scenario investigates the effect of the application of the model with a weaker shock wave compared to the first three scenarios. The best-fitting parameters for each of the following scenarios are presented in Table~\ref{tab: Parameter values all scenarios} and will be described in detail in the following sections. A comparison of the resulting radio spectral indices between 0.325 and 3.0 GHz for the four cases is presented in Table~\ref{tab: Calculated spectral indices}.

\begin{table*}
	\caption[Parameter values]{Best-fitting values of the phase parameters used in each scenario with the calculated cumulative compression ratios, magnetic field values, magnetic energy densities and characteristic momenta. The parameter values for phases 0 to 2 are the same in all the scenarios. The values for phases 3 and 4 are listed for each scenario.}
	\label{tab: Parameter values all scenarios}
	\centering
	\begin{tabular}{ccccccccc}
		\hline
		\rule{0pt}{2.2ex} {Phase} & {$\Delta t_i$} & {$\tau_i$} & {$C_{i-1 \, i}$} & {$b_i$} & $C_{0\,i}$ & $B_i$ &
		$u_{\text{B}\,i}$ & $p_{*0\,i}$
		\\
		& Myr & Myr & & & & $\muup \text{G}$ & $\text{eV cm}^{-3}$ &
		\\
		\hline
		All scenarios & & & & & & & &
		\\
		\rule{0pt}{2.2ex}0 & 0 & 15.0 & 1 & 1.8 & 1.0  & 1.90 & 0.0899 & $\infty$ 
		\\
		1 & 3.46 & 10.0 & 0.7 & 1.2 & 0.70 &  1.50 & 0.0559 & $303\,000$
		\\
		2 & 54 & $\infty$ & 1 & 0 & 0.70 & 1.50 & 0.0559 & $22\,100$
		\\
		\hline
		Scenario 1 & & & & & & & &
		\\
		3 & 2.5 & $-6.84$ & 2.48 & 2 & 1.74 & 2.75 & 0.188 & $21\,200$
		\\
		4 & 28 & $\infty$ & 1 & 0 & 1.74 & 2.75 & 0.188 & $11\,900$
		\\
		\hline
		Scenario 2 & & & & & & & &
		\\
		3 & 2.5 & $-6.84$ & 2.48 & 2 & 1.74 & 2.75 & 0.188 & $21\,200$
		\\
		4 & 50 & $\infty$ & 1 & 0 & 1.74 & 2.75 & 0.188 & $8\,860$
		\\
		\hline
		Scenario 3 & & & & & & & &
		\\
		3 & 2.5 & $-6.84$ & 2.48 & 2 & 1.74 & 2.75 & 0.188 & $21\,200$
		\\
		4 & 22 & $88.0$ & 0.8 & 1 & 1.39 & 2.37 & 0.140 & $13\,200$
		\\
		\hline
		Scenario 4 & & & & & & & &
		\\
		3 & 0.10 & $-0.948$ & 1.25 & 2 & 0.875 & 1.74 & 0.0752 & $22\,000$
		\\
		4 & 20 & $\infty$ & 1 & 0 & 0.875 & 1.74 & 0.0752 & $15\,900$
		\\
		\hline
		
	\end{tabular}
\end{table*}

\begin{figure*}
	\subfloat{
		\includegraphics[width=0.5\linewidth]{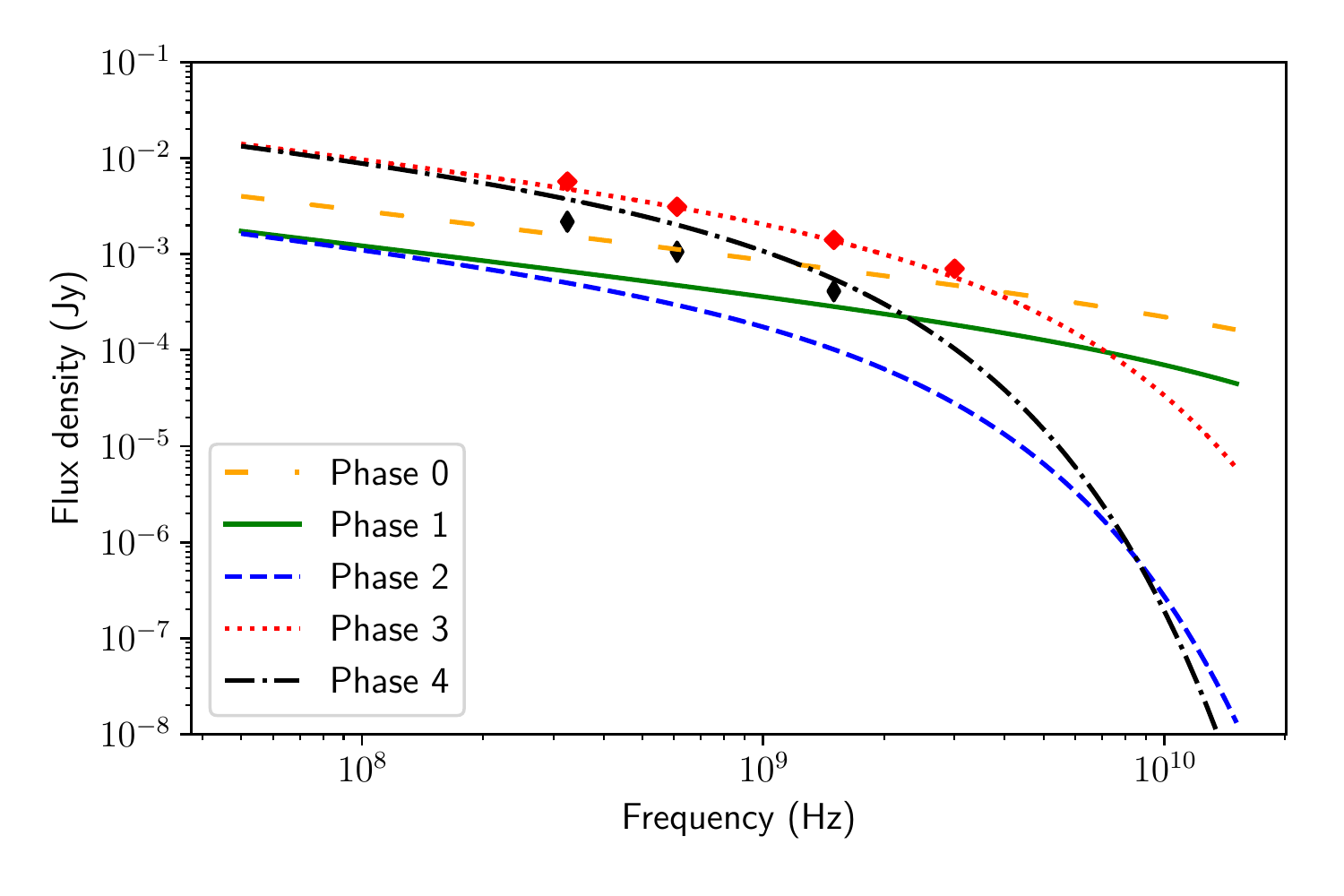}
	}
	\subfloat{
		\includegraphics[width=0.5\linewidth]{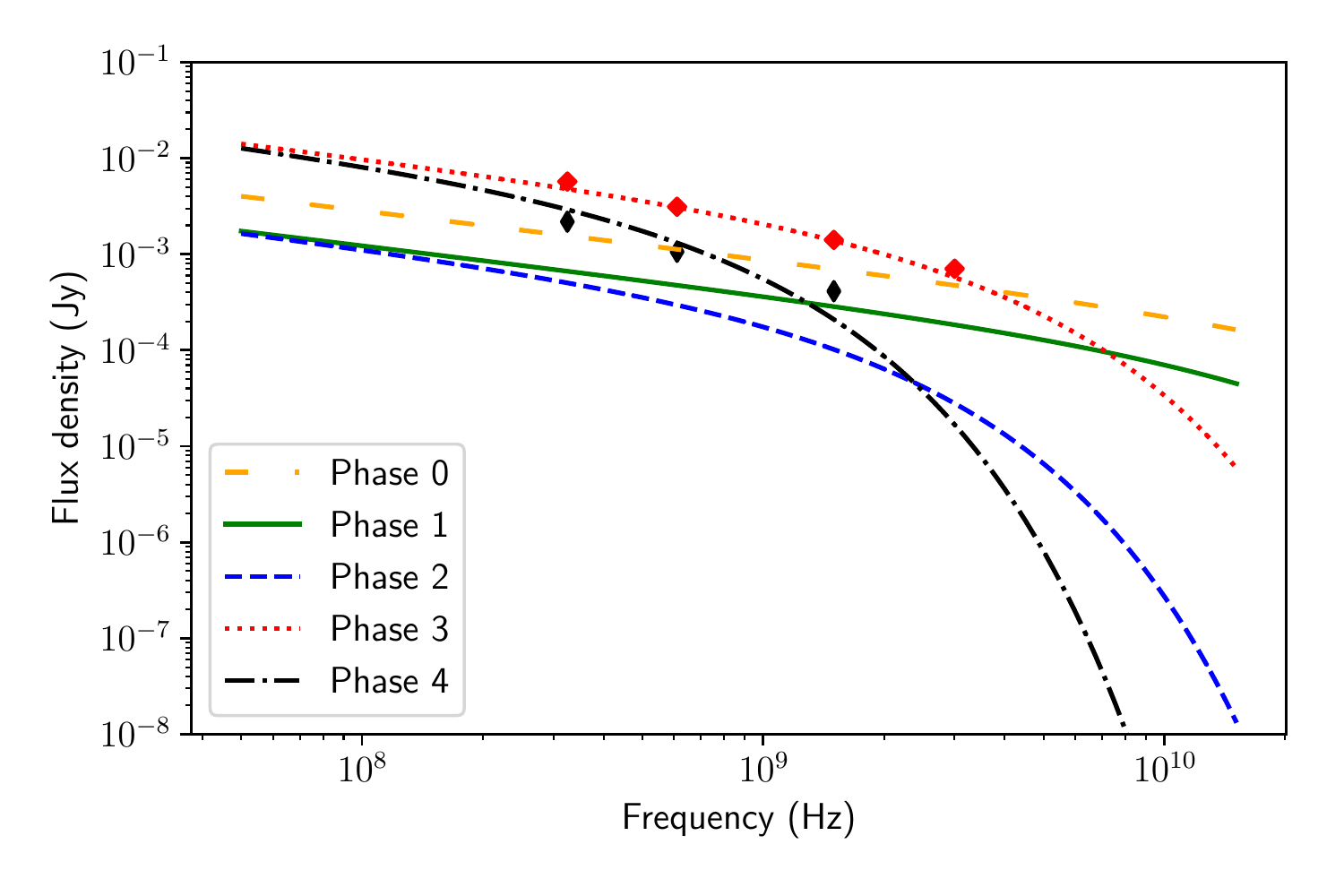}
	}
	\newline	
	\subfloat{
		\includegraphics[width=0.5\linewidth]{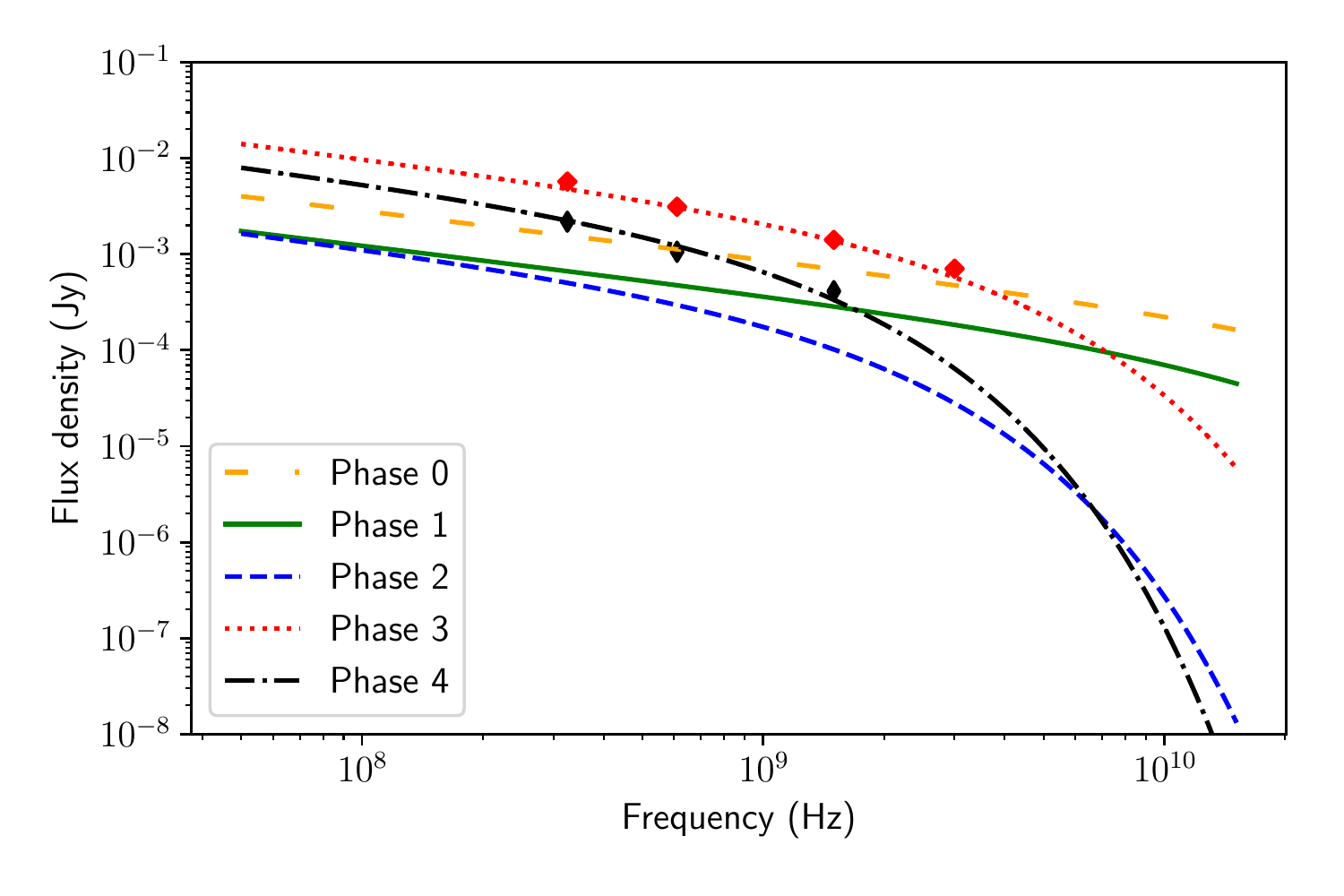}
	}
	\subfloat{
		\includegraphics[width=0.5\linewidth]{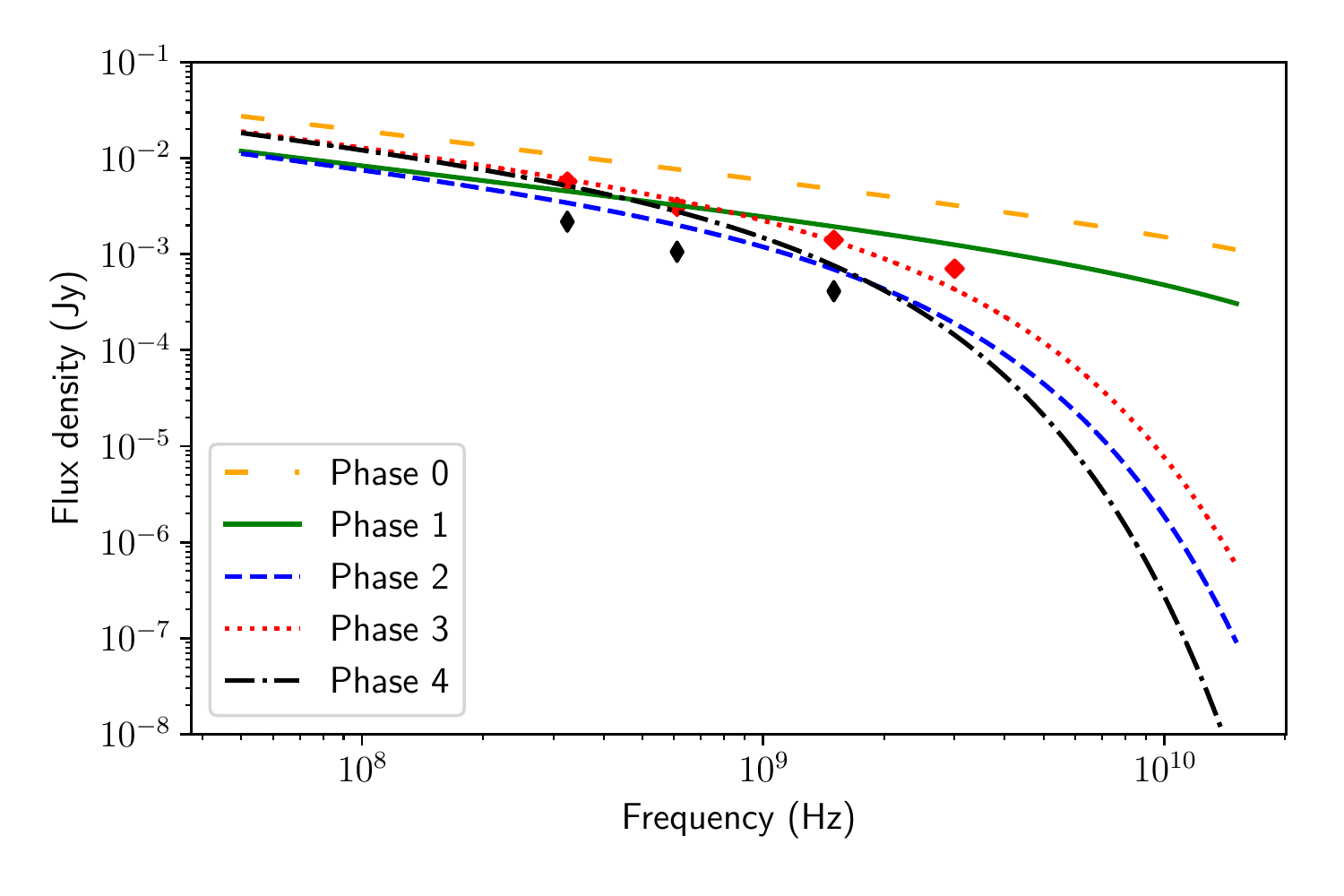}
	}
	\caption{Resulting flux densities. The measured flux densities at 80 kpc and 140 kpc are shown in the red and black point series, respectively. \textit{Top left:} Results from the scenario 1 modelling, where the electron population was injected by the radio galaxy and the evolution of the electron spectrum is studied through all the phases of the adiabatic model. \textit{Top right:} Results from the scenario 2 modelling where the duration of phase 4 was increased to produce a steeper spectrum behind the shock front. \textit{Bottom left:} Results from the modelling in scenario 3, where the radio cocoon continues to expand during the final phase of the model. \textit{Bottom right:} Results from the modelling in scenario 4, where the electron population was injected by the radio galaxy and the adiabatic model is applied with a weaker shock value of $\mathcal{M} = 1.13$.}
	\label{fig: Flux densities for all scenarios}
\end{figure*}

\subsection{Scenario 1: Application of the adiabatic compression model}
\label{sec: Scenario 1: Application of the full adiabatic model}
This scenario investigates the application of the adiabatic compression model to the relic according to the formalism presented above. Since the value of $u_{\text{B\,0}}$ is not known in this case, the value of the compression ratio of phase 1 is taken as a free parameter. However, since $u_{\text{B}\,0}$ is related to $u_{\text{B}\,1}$ by the factor $(C_{0\,1})^{4/3}$, the value of $C_{0\,1}$ allows us to constrain the value of $u_{\text{B\,0}}$ which affects the value of the spectral index of phase 0. Furthermore, since $C_{0\,2}$ depends on the $C_{0\,1}$, the value of $C_{0\,1}$ also affects the spectral index of phase 2. Thus, $C_{0\,1}$ was chosen to produce the best-fitting values of the spectral indices of phase 0 and phase 2; the resulting value was $C_{0\,1} = 0.7$. The resulting duration of phase 1 was then calculated to be $\Delta t_1 = 3.46$ Myr. The durations of phases 2 and 3 that best reproduce the spectral indices at the end of the radio tail and at the shock front were found to be 54 Myr and 2.5 Myr, respectively, with the resulting value of $\tau_3$ calculated to be $-6.84$ Myr. Lastly, the duration of phase 4 that reproduces the spectral index at 140 kpc is 28 Myr. The best-fitting model values are listed in Table~\ref{tab: Parameter values all scenarios} and the resulting radio spectra are shown in the top left panel of Fig.~\ref{fig: Flux densities for all scenarios}.

This scenario is successful in reproducing the measured flux density at the position of the shock front. The calculated spectrum at the end of phase 3, shown by the dotted red curve in the top left panel of Fig.~\ref{fig: Flux densities for all scenarios}, clearly follows the shape of the measured spectrum at 80 kpc. This is confirmed by considering the radio spectral index, as the calculated value of $-0.954$ matches the measured value of $-0.95$. Furthermore, the curve also fits the measured level of the flux density at this position. This is expected since the normalization of the electron population was chosen by fitting the flux density at this position. The observations of the relic, as shown in the radio maps in \citet{van_Weeren_et_al_2017}, indicate that the emission of the galaxy is weaker than the peak emission of the relic. Thus, since the calculated spectrum at the end of phase 0 (shown in the solid orange curve) is below the peak emission of the relic for frequencies below 3.0 GHz, this value of the normalization is reasonable.

The radio spectrum at the end of phase 4, shown by the dot-dashed black curve in the top left panel of Fig.~\ref{fig: Flux densities for all scenarios}, lies above the measured radio spectrum at 140 kpc even though it has the same spectral index between 0.325 and 3.0 GHz. Increasing the duration of phase 4, with the aim to obtain a lower value of the flux, was found to produce a spectral index that is steeper than the observed one. This indicates that further investigation of the radio spectrum behind the shock front is required. The next two scenarios investigate this problem.

\subsection{Scenario 2: Investigation of a steeper spectrum behind the shock front}
The previous scenario demonstrates that although there seems to be some difficulty in recovering the flux density and the measured radio spectral index at 140 kpc, the adiabatic model is able to reproduce the spectral shape of the radio emission at the shock front. In this and the next section, we investigate two different methods to try to resolve the difficulty of the emission behind the shock front. In this scenario, the radio spectrum behind the shock front was allowed to be steeper than the measured spectral index at 140 kpc and the calculated spectrum was matched to the flux density rather than the spectral shape.

The values of all the phase parameters before phase 4 were chosen in the same way as in the previous case. The duration of phase 4 was chosen as the value that best reproduced the flux density at 140 kpc. The best-fitting value was found to be 50 Myr. The phase parameters are listed in Table~\ref{tab: Parameter values all scenarios} and the resulting radio spectra are shown in the top right panel of Fig.~\ref{fig: Flux densities for all scenarios}.

The longer duration of phase 4 in this scenario compared to that in scenario 1 in section \ref{sec: Scenario 1: Application of the full adiabatic model}, produces a lower flux density. This is expected since the longer the duration of this phase the more the radiative losses and so the lower the flux density. As shown by the dot-dashed black curve in the top right panel of Fig.~\ref{fig: Flux densities for all scenarios}, the calculated spectrum fits the measured flux density at 325 and 610 MHz but lies below the measured flux density at 1.5 GHz, implying that the correct spectral shape at high frequencies is not reproduced in this scenario.

This effect, that the calculated spectrum steepens more than the observed spectrum at high frequencies, could indicate that other mechanisms, such as turbulence \citep{Fujita_et_al_2015}, are effective behind the shock front and continue to energize the electrons in this region. Another possibility is that the radio cocoon starts to expand slightly after the shock passage. This would result in a weaker magnetic field which could make the losses due to the expansion more important than the radiative losses, possibly providing a temporarily flatter spectrum, compared to where the losses are only radiative. The effect of such a mechanism is investigated in section \ref{sec: Scenario 3: Investigation of an expansion phase behind the shock front}.

\subsection{Scenario 3: Investigation of an expansion phase behind the shock front}
\label{sec: Scenario 3: Investigation of an expansion phase behind the shock front}
The previous scenario demonstrates that it is difficult to match the flux density behind the shock front at both the low frequency and high frequency at the same time by considering only radiative energy losses. This scenario investigates the effects of an expansion of the radio cocoon following the compression caused by the shock passage.
\citet{Ensslin_&_Gopal-Krishna_2001} assumed that there would be no expansion for this phase, however, if the blob after the shock passage is slightly over-pressured compared to the surrounding medium it is possible that there could be some expansion. In this scenario we assume that the rate of expansion behind the shock front is lower than the expansion rate in phase 1.

Again, the values of the phase parameters until phase 4 were chosen in the same way as in scenario 1 in section \ref{sec: Scenario 1: Application of the full adiabatic model}. The parameters of phase 4 were chosen to give the best-fitting results for both the flux density and the spectral index without working out a physical model of the expansion. The compression ratio, $C_{3\,4}$ was chosen to be $0.8$ since this represents an expansion of the cocoon but is still closer to one compared to the value of 0.7 assumed for phase 1, corresponding to a slower expansion rate. The exponent of the expansion, $b_4$, was chosen to be 1 which would correspond to an expansion that is not as rapid as the expansion during phase 1. The duration of the phase, $\Delta t_4$, is still a free parameter and its value is chosen to give the best-fitting spectral index. Finally, the value of $\tau_4$ is calculated from the values of $C_{3\,4}$, $b_4$ and $\Delta t_4$ according to eqn \eqref{eqn: Phase parameters}. The best-fitting parameter values are listed in Table~\ref{tab: Parameter values all scenarios} and the resulting radio spectra are shown in the bottom left panel of Fig.~\ref{fig: Flux densities for all scenarios}.

As shown by the dot-dashed black curve in the bottom left panel of Fig.~\ref{fig: Flux densities for all scenarios}, this assumption of an expansion behind the shock front can fit the shape of the radio spectrum behind the shock front at all the measured frequencies. This scenario was investigated by finding the best-fitting values of the parameters that reproduce the flux density across all the measured frequencies. However, further modelling of the physical process which would cause the cocoon to expand after the shock passage is needed in order to be able to further investigate this scenario and whether it is plausible as a model.

\subsection{Scenario 4: Investigation of the adiabatic model with a smaller shock Mach number}
As mentioned in section \ref{sec: The case of Abell 3411-3412}, the Mach number of the shock in this region of the galaxy cluster is difficult to determine with much certainty. The value of the shock Mach number used in the modelling above is $\mathcal{M} = 1.7$, and, although this value is somewhat smaller than the value of $\mathcal{M} = 1.9$ which \citet{van_Weeren_et_al_2017} found using the DSA-type re-acceleration model, the value of the Mach number could be even smaller. The recent determination of the shock Mach number obtained a value of $\mathcal{M} = 1.13$ \citep{Andrade-Santos_et_al_2019}. This section investigates the application of the adiabatic model to the radio relic using this smaller value of the shock Mach number.

Since the flux density of the galaxy is not reported, in the previous scenarios the constant $\tilde{N_0}$ in the initial electron spectrum was estimated by matching the calculated flux density at the end of phase 3 with the peak flux density of the relic. In this scenario, the smaller shock Mach number will result in a smaller compression which will result in a smaller level of the flux density in phases 3 and 4. Thus, the value of $\tilde{N_0}$ needs to be estimated again for this scenario. The best-fitting value that matched the flux density at the end of phase 3 to the peak flux density of the relic was found to be $\tilde{N_0} = 1.5 \times 10^{62}$.

The value of the compression ratio of phase 3 calculated for the smaller Mach number was found to be  $C_{2\,3} = 1.25$, which gives a post-shock magnetic field of $B_3 = 1.74 \text{ } \muup \text{G}$. The values of $\Delta t_3$ and $\Delta t_4$ that gave best-fitting values of the spectral indices at the end of phases 3 and 4 were found to be $\Delta t_3 = 0.1$ Myr and $\Delta t_4 = 20$ Myr. The resulting value of $\tau_3$ was $-0.948$ Myr. All the other phase parameters were kept the same as in scenario 1. The values of the phase parameters and the resulting magnetic field values and characteristic momenta are listed in Table~\ref{tab: Parameter values all scenarios}. The resulting radio spectra are shown in the bottom right panel of Fig.~\ref{fig: Flux densities for all scenarios}.

As shown by the dotted red curve in the bottom right panel of Fig.~\ref{fig: Flux densities for all scenarios} the spectrum at the end of phase 3 is steeper than the observed spectrum. Furthermore, the level of the flux density at the end of phase 3 is lower than the calculated level of the flux density at the end of phase 0 and 1. As mentioned in section \ref{sec: Scenario 1: Application of the full adiabatic model} the radio maps indicate that the emission of the galaxy is weaker than the peak emission of the relic and this is not reproduced in this modelling. Although the spectral shape at the end of phase 4 (shown in the dot-dashed black curve) does seem to match the spectral shape of the flux density behind the shock front, the level of the calculated flux density is greater than the measured flux density behind the shock front.

These results would seem to indicate that at this level of the shock Mach number, the compression is insufficient to re-energize the electrons to high enough energies so that the spectral shape at the shock front is reproduced or that the emission at the shock front is greater than the emission from the galaxy. 

\begin{table}
	\caption[Calculated spectral indices for the first four scenarios]{Comparison of the calculated spectral indices between 0.325 and 3.0 GHz at the end of each phase of the adiabatic model for the first four scenarios considered.}
	\label{tab: Calculated spectral indices}
	\centering
	\begin{tabular}{ccccc}
		\hline
		\rule{0pt}{2.2ex}{Phase} & \multicolumn{4}{c} {{Spectral index}}
		\\
		&  {Case 1} & {Case 2} & {Case 3} & {Case 4}
		\\
		\hline
		\rule{0pt}{2.2ex}0 & $-0.535$ & $-0.535$ & $-0.535$ & $-0.535$
		\\
		1 & $-0.578$ & $-0.578$ & $-0.578$ & $-0.578$
		\\
		2 & $-1.30$ & $-1.30$ & $-1.30$ & $-1.30$
		\\
		3 & $-0.954$ & $-0.954$ & $-0.954$ & $-1.19$
		\\
		4 & $-1.61$ & $-2.32$ & $-1.60$ & $-1.61$
		\\
		\hline
	\end{tabular}
\end{table}

\section{Conclusions}
This paper has looked at four scenarios in which the adiabatic compression model was applied to the north-eastern component of the radio relic that is hosted in the merging galaxy cluster A3411-3412. In the first three scenarios, a shock strength of $\mathcal{M}=1.7$ was used while in the last scenario a much weaker shock of $\mathcal{M} = 1.13$ was used, and the free model parameters were chosen so that the observed spectral index at the end of each phase matched the spectral index at several points across the relic. The first scenario simply applied the model, as given in the formalism presented in \citet{Ensslin_&_Gopal-Krishna_2001}, to the relic and was successful in reproducing the spectral shape at and behind the shock front. However, behind the shock front the calculated flux density was greater than the observed one. Scenarios 2 and 3 investigated different effects behind the shock front to try to reproduce the spectral shape and the flux density in this region. Scenario 2 showed that it is difficult to reproduce the flux density behind the shock at all three frequencies, 0.325, 0.610 and 1.5 GHz, by considering only radiative losses which could indicate that other mechanisms, such as turbulence or the expansion of the cocoon, take place in this region. Scenario 3, which investigated an expansion phase behind the shock front based on best-fitting parameters, was able to reproduce the spectral shape and the flux density, however, further modelling of such a scenario is needed in order to investigate this mechanism more fully. Finally, scenario 4 investigated an application of the model at a smaller Mach number of 1.13, according to more recent results. This smaller Mach number would result in a smaller compression caused by the shock wave and the results of this scenario seem to suggest that the adiabatic compression model has difficulty in reproducing the observed spectral index at this smaller Mach number.

These results show that the adiabatic compression scenario can reproduce the observed spectrum of the R1 region of the radio relic in the cluster A3411-3412 at the relic peak and also in the region behind the relic if a further expansion of the blob takes place after the shock passage. Post-shock turbulence could also describe the spectral shape behind the shock front. In this case the electrons would also gain energy by turbulent re-acceleration \citep{Fujita_et_al_2015, Fujita_et_al_2016} so that the re-energization would no longer be due, purely, to adiabatic compression. We also note that these results were produced using a value of the shock strength that was smaller than the value predicted by the DSA-type re-acceleration model and closer to the one indicated by the X-ray measurements reported in \citet{van_Weeren_et_al_2017}. However, at the smaller Mach number determined by \citet{Andrade-Santos_et_al_2019}, the adiabatic model does not seem to be able to recover the observed spectral indices across the relic, but, as discussed, there is some uncertainty in determining the value of the shock Mach number in this region of the cluster.

As previously mentioned the, size of the R1 region of the relic in A3411-3412 is quite small compared to the extension of the whole radio structure and to the typical extension of relics. It is therefore possible that if the adiabatic compression scenario is the correct one for the R1 region, this source should not be referred to as a radio gischt/radio shock (according to the respective classifications used by \cite{Kempner_et_al_2004} and \cite{van_Weeren_et_al_2019}), but as re-energized AGN-plasma. The rest of the whole structure might have a different origin and might be produced by the same interaction between the radio jet of the galaxy and the shock front that can produce a complex structure depending on the time elapsed from the interaction and the geometry of the system \citep{Pfrommer_&_Jones_2011}.

The comparison of the calculated flux densities to the measured flux densities is somewhat inconclusive due to a lack of some observational data. The normalization constant of the electron spectrum of the radio galaxy was only estimated based on the peak flux density of the relic. Further observations of the galaxy, particularly obtaining measurements of the flux density of the galaxy and the magnetic field within the galaxy, are needed to determine this constant more accurately. The magnetic field of the galaxy will also constrain the value of the compression ratio of phase 1. Furthermore, observations that provide a better understanding of the structure of the relic and the flux densities and spectral indices across the relic are needed for a comparison with the fluxes predicted by the adiabatic model. Since the duration of phase 3 is related to the plasma volume and pre-shock flow velocity, deeper observations of the radio tail that can provide an estimate these quantities would constrain the value of the duration of phase 3. These improvements would enable a more accurate comparison of the calculated flux densities to the measured flux densities which would enable this application of the adiabatic compression model to be tested further.

Additionally, since the adiabatic model predicts that the galaxy emission should start to dominate at frequencies greater than 3 GHz, observations of the galaxy-relic system at higher frequencies could test the predictions of the adiabatic compression model. MeerKAT and the upcoming SKA will be very important for such observations given their sensitivity and angular resolution that should allow for the detection of the radio emission, as well as the measurement of the spectrum at several frequencies, even in low-brightness regions such as the end of the radio tail.

\section*{Acknowledgements}
The authors would like to thank the referee for useful comments. CB and PM acknowledge support from the DST/NRF SKA post-graduate bursary initiative.

\section*{Data availability}
No new data were generated or analysed in support of this research.


\bibliographystyle{mnras}
\bibliography{BibData}

\bsp	
\label{lastpage}
\end{document}